\documentclass{article}
\usepackage[margin=1in]{geometry}
\usepackage[utf8]{inputenc}
\usepackage{mathtools}
\usepackage{amsmath}
\usepackage{amssymb}
\usepackage[T1]{fontenc}
\usepackage{longtable}
\usepackage{topcapt}
\usepackage{graphicx}
\usepackage[parfill]{parskip}
\usepackage{tikz}
\usepackage{booktabs}
\usepackage{array}
\usepackage{multirow}
\usepackage{wrapfig}
\usepackage{float}
\usepackage{colortbl}
\usepackage{pdflscape}
\usepackage{caption}
\usepackage{setspace}
\usepackage[normalem]{ulem}
\usepackage{makecell}
\usepackage{authblk}
\DeclareMathOperator*{\argmax}{arg\,max}

\restylefloat{table}
\usetikzlibrary{positioning}

\usepackage[backend = biber, 
            sorting=none,
            style=nejm,
            citestyle = numeric]{biblatex}
\title{Causal mediation analysis with mediator values below an assay limit}
\author[1]{Ariel Chernofsky}
\author[2]{Ronald Bosch}
\author[3]{Judith J. Lok}
\affil[1]{Department of Biostatistics, Boston University School of Public Health, MA, USA}
\affil[2]{Center for Biostatistics in AIDS Research, Harvard T.H. Chan School of Public Health, Boston, MA}
\affil[3]{Department of Mathematics and Statistics, Boston University, MA, USA}
\date{\today}

\begin{document}

\doublespacing

\maketitle

\abstract{Causal indirect and direct effects provide an interpretable method for decomposing the total effect of an exposure on an outcome into the effect through a mediator and the effect through all other pathways. When the mediator is a biomarker, values can be subject to an assay lower limit. The mediator is affected by the treatment and is a putative cause of the outcome, so the assay lower limit presents a compounded problem in mediation analysis. We propose three approaches to estimate indirect and direct effects with a mediator subject to an assay limit: 1. extrapolation 2. numerical optimization and integration of the observed likelihood and 3. the Monte Carlo Expectation Maximization (MCEM) algorithm. Since the described methods solely rely on the so-called Mediation Formula, they apply to most approaches to causal mediation analysis: natural, separable, and organic indirect and direct effects. A simulation study compares the estimation approaches to imputing with half the assay limit. Using HIV interruption study data from the AIDS Clinical Trials Group described in [Li et al. 2016, AIDS; Lok \& Bosch 2021, Epidemiology], we illustrate our methods by estimating the organic/pure indirect effect of a hypothetical HIV curative treatment on viral suppression mediated by two HIV persistence measures: cell-associated HIV-RNA (N = 124) and single copy plasma HIV-RNA (N = 96).}

\newpage

\section{Introduction}\label{sec:intro}

\begin{figure}[h]
\captionsetup{width=0.8\textwidth}
\caption{\small Causal diagram with assumed data structure. Arrows from one node to another indicate a causal relationship.}
\begin{center}
{\begin{tikzpicture}[%
	->,
	>=stealth,
	node distance=2cm,
	pil/.style={
		->,
		thick,
		shorten =6 pt,}
	]
	\node (1) {M: mediator};
	\node[left=of 1] (2) {A: treatment};
	\node[right=of 1] (3) {Y: outcome};
	\node[below= of 1] (4) {C: common causes};
	\draw [->] (1.east) -- (3.west);
	\draw [->] (2) to [out=30, in=150] (3);
	\draw [->] (2.east) to (1);
	\draw [->] (4) to (1);
	\draw [->] (4) to (3);
\end{tikzpicture}}
\end{center}
\label{fig:dag1}
\end{figure}
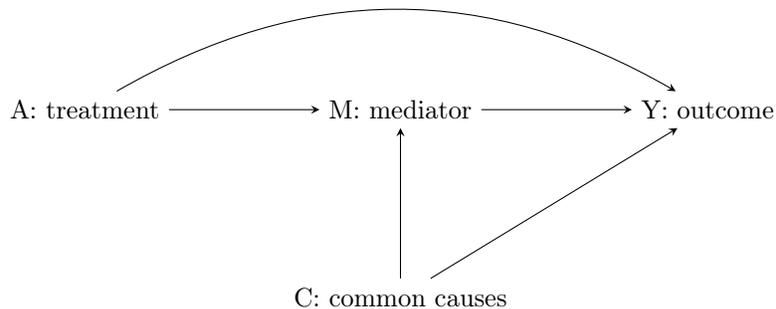

Recent causality research extends the notion of a causal effect beyond the "simple" randomized controlled trial (RCT). While RCTs can provide a result with a causal interpretation, their findings are often a "black box view of causality" \cite{imai2010a}. For example, consider an RCT designed to study the effect of a new drug that is hypothesized to target a pathway that may stall disease progression. If designed correctly, a successful RCT can establish causality between the new treatment and disease progression, but the initial biological hypothesis remains unanswered. Mediation analysis answers questions about how and why a drug affects disease progression. Baron and Kenny \cite{baron1986} established a foundation for mediation, which decomposes the total effect of an exposure on an outcome into 1. the indirect path from an exposure to the outcome through a mediator variable and 2. the direct pathway from the exposure to the outcome through all other pathways. The indirect and direct effects can be visualized through the causal diagram in Figure \ref{fig:dag1}. Let $A$ be a binary randomized treatment, $Y$ the outcome of interest, $M$ a mediator measured between $A$ and $Y$. Since $M$ is not randomized, pre-treatment mediator-outcome common causes $C$ must be adjusted for \cite{robins1992, pearl2001}. 
Arrows directed from one node to another imply a causal relationship. Since $A$ is assumed randomized, there are no common causes of $A$ and $Y$, but the randomization assumption can be relaxed \cite{lok2016}.  

Pearl \cite{pearl2001} and Robins and Greenland \cite{robins1992} established a causal counterfactual framework for causal mediation known as the natural indirect and direct effects. They use counterfactual quantities: $M^{(a)}$ and $Y^{(a)}$, mediator and outcome with treatment $A$ set to $a$ (0 or 1), and  $Y^{(a, m)}$, the outcome with treatment $A$ set to $a$ (0 or 1) and mediator $M$ set to value $m$. The natural indirect and direct effects rely on $Y^{(1, M^{0})}$, the outcome under $a = 1$ but with the mediator taking its value under $a = 0$: a so called cross-worlds term that exists in two mutually exclusive realities. The natural indirect effect $$\mathbb{E}[Y^{(1)}] - \mathbb{E}[Y^{(1, M^{0})}]$$ evaluates the expectation of an outcome $Y$ while varying mediator $M$ from $M^{(1)}$, its value under $a = 1$, to $M^{(0)}$, its value under $a = 0$, while setting the value of the treatment $A$ to $a = 1$. The natural direct effect $$\mathbb{E}[Y^{(1, M^{0})}] - \mathbb{E}[Y^{(0)}]$$ evaluates the expectation of the outcome $Y$ under varied treatment $A$ statuses $a = 1$ vs. $a = 0$ with  $M$ set to $M^{(0)}$, the value $M$ would attain under treatment status $a = 0$. Identifiability of natural indirect and direct effects requires restrictive assumptions on $Y^{(1, M^{0})}$ that are not verifiable but yield the Mediation Formula \cite{pearl2001}
\begin{equation}\label{eqn:natmedform}
    \mathbb{E}[Y^{(1, M^{0})}] = \int_{m, c} \mathbb{E}[Y\mid M = m, A = 1, C = c]f_{M \mid A=0, C = c}(m) f_C(c) dmdc,
\end{equation}
which identifies the cross-worlds term from observable and estimable quantities.

Lok \cite{lok2016} proposed organic indirect and direct effects, which avoid  cross-worlds counterfactuals and relax identifiability assumptions. They generalized this approach to the organic indirect and direct effects relative to a specific value $A = a$ \cite{lok2021}. Let $I$ be an intervention on the mediator. Denote $M^{(a, I=1)}$ and $Y^{(a, I=1)}$ as the mediator and outcome respectively, under $A = a$ and combined with intervention $I$ on the mediator. Organic indirect and direct effects relative to $a = 1$ generalize the natural indirect and direct effects with relaxed identification assumptions. Here we mainly consider organic indirect and direct effects relative to $a = 0$. $I$ is said to be an organic intervention relative to $a = 0$ and $C$ if 
\begin{equation}\label{eqn:i1}
    M^{(0, I = 1)} \mid  C = c \quad \sim \quad M^{(1)} \mid C = c
\end{equation}
and 
\begin{equation}\label{eqn:i2}
    Y^{(0, I = 1)} \mid M^{(0, I = 1)} = m, C = c \quad \sim \quad Y^{(0)} \mid M^{(0)} = m, C = c.
\end{equation}
In contrast to natural indirect and direct effects, which set the value of the mediator, (\ref{eqn:i1}) $I$ "holds the mediator at its distribution under treatment" \cite{lok2021}. Equation (\ref{eqn:i2}) requires that $I$ is only associated with $Y$ through its effect on the mediator. The organic indirect and direct effects relative to $a = 0$ and $C$ are defined as
    \begin{equation} \label{neworgdirect}
        \text{Indirect effect: }\mathbb{E}[Y^{(0, I = 1)}] - \mathbb{E}[Y^{(0)}],    
    \end{equation}
    and
    \begin{equation}\label{neworgindirec}
        \text{Direct effect: }\mathbb{E}[Y^{(1)}] - \mathbb{E}[Y^{(0, I = 1)}].
    \end{equation}
Assuming a randomized treatment and consistency, \cite{lok2021} showed that the Mediation Formula (\ref{eqn:natmedform}) holds for organic interventions:
\begin{equation}\label{newmedform}
    \mathbb{E}[Y^{(0, I = 1)}] = \int_{m, c} \mathbb{E}[Y\mid M = m, A = 0, C = c]f_{M \mid A=1, C = c}(m) f_C(c) dmdc.
\end{equation}
Compared to the Mediation Formula relative to $a = 1$ and $C$, the Mediation Formula relative to $a = 0$ and $C$ depends on outcome data exclusively from untreated participants, which has two key advantages \cite{lok2021}. First, the mediation formula  easily accommodates mediator-treatment interactions in the outcome model. Second, if the effect of treatment on the mediator distribution is known, the indirect effect can be estimated in the absence of outcome data under treatment. Regardless of the choice of causal indirect and direct effects, the result is a Mediation Formula that informs estimation of causal mediation. The estimation methods described here rely on the Mediation formula, so they apply to natural, pure, organic, and separable indirect and direct effects. 

If the mediator is measured by an assay, the measurements can sometimes fall below the assay limit. Since the mediator is an effect of the treatment and a cause of the outcome, the problem of an assay limit, sometimes referred to as left censoring, is compounded in mediation analysis. As a motivating example, suppose we are interested in estimating the effect of a new curative HIV treatment on viral suppression, mediated by its effect on viral persistence.  However, the measurement of viral persistence falls below the assay lower limit for many participants, leading to a left censored mediator. The problem of left censoring has been addressed for scenarios when the censored variable is the outcome \cite{lyles2001} or the predictor \cite{lynn2001}, but these methods have not been applied to causal mediation, where the censored mediator is both an outcome and a predictor. 

In this paper we develop methods for addressing the challenge of mediators with assay limits when estimating causal indirect and direct effects. The paper is organized as follows. Section \ref{sec:semip} describes a semi-parametric estimator for indirect and direct effects identified by the so-called mediation formula. Section \ref{sec:semip_al} focuses on estimating indirect and direct effects for mediators with assay limits and proposes three estimation methods that leverage observed data to account for the informatively missing mediator: 1. mediator model extrapolation 2. numerical integration and optimization of the observed data likelihood function and 3. the Monte Carlo Expectation-Maximization (MCEM) algorithm. Section \ref{sec:sim} presents the results of a simulation study comparing these three methods to a simple imputation technique replacing censored values with assay limit divided by 2, and evaluating the performance of the methods for a small and large sample size with many mediator values below the assay limit. Section \ref{sec:dataapp} illustrates the proposed methods with HIV treatment interruption data from the AIDS Clinical Trials Group \cite{li2016} and estimates the indirect effect of a curative HIV treatment on viral rebound through week 4 mediated through two viral persistence measures. 

\section{Semiparametric estimators for the organic direct and indirect effects relative to a = 0 }\label{sec:semip}

Baron and Kenny \cite{baron1986} introduced the product method to estimate the indirect and direct effects. The product method fits the following models: 
\begin{equation}\label{eqn:bk1}
    \mathbb{E}[Y \mid A, M, C] = \beta_0 + \beta_1 A + \beta_2 M + \beta_3^\top C,
\end{equation}
and
\begin{equation}\label{eqn:bk2}
    \mathbb{E}[M \mid A, C] = \alpha_0 + \alpha_1 A + \alpha_2 ^\top C.
\end{equation}
The direct effect is $\beta_1$, the adjusted effect of treatment on the outcome, and the indirect effect is $\beta_2 \times \alpha_1,$ the product of the adjusted effect of treatment on the mediator and the adjusted effect of the mediator on the outcome. The Ordinary Least Squares (OLS) method estimates of $\alpha_1$, $\beta_1$, and $\beta_2$ induce consistent and asymptomatically normal estimators of the indirect and direct effects. If $(M_i, Y_i)_{i = 1,...,N}$ are continuous independent and identically distributed random variables then the estimates are also efficient. Under the organic intervention approach to mediation described in Section \ref{sec:semip_al} and correct model specification of (\ref{eqn:bk1}) and (\ref{eqn:bk2}), the product method yields the organic indirect and direct effects. In fact, for organic \cite{lok2021} and pure \cite{robins1992} indirect and direct effects relative to $a = 0$, the product method still holds if the outcome model has a treatment-mediator interaction between  $A$ and $M$. However, non-linear outcome models and the difficulty of correct model specification limit the applicability of the product method. Imai et al. \cite{imai2010a} suggest a simulation approach that relies on mediator model assumptions that are not necessary. An alternative approach models the mediation formula to directly estimate the counterfactual expectations \cite{lok2021}. This method requires only specification of an outcome model under no treatment to estimate $\mathbb{E}[Y^{(0, I = 1)}]$. For observed randomized treatment data $(Y_i, M_i, A_i, C_i)_{i = 1,...,N},$ an estimate for $\mathbb{E}[Y^{(0, I = 1)}]$ is: $$\hat{\mathbb{E}}[Y^{(0, I = 1)}] = \frac{1}{\sum A_i}\sum_{i:A_i = 1}\mathbb{\hat{E}}[Y\mid M_i = m, A_i = 0, C_i = c].$$
To estimate $\mathbb{\hat{E}}[Y\mid M_i = m, A_i = 0, C_i = c]$ we specify an outcome model for untreated subjects and obtain predicted values for treated subjects. The remaining terms $\mathbb{E}[Y^{(0)}]$ and $\mathbb{E}[Y^{(1)}]$ are estimated without parametric assumptions using sample averages of the outcome for untreated and treated observations, respectively, yielding a semi-parametric estimator for the indirect and direct effects. The estimation procedure for any outcome $Y$ type can be summarized by the following three steps:
\begin{enumerate}
    \item Fit a generalized linear model with link function $g(\cdot)$ on untreated subjects:
$$g(\mathbb{E}[Y_i \mid A = 0, M_i, C_i]) = \beta_0 + \beta_2 M_i + \beta_3^\top C_i,$$
to obtain $\hat{\beta}.$
    \item Estimate $$\hat{\mathbb{E}}[Y_i \mid A = 0, M_i, C_i] = g^{-1}(\hat{\beta}_0 + \hat{\beta}_2 M_i + \hat{\beta}_3^\top C_i),$$ for all $i = 1,...,N$ with $A_i = 1.$
    \item Estimate the organic direct effect as
$$\hat{\mathbb{E}}[Y^{(1)}] - \hat{\mathbb{E}}[Y^{(0, I = 1)}] = \frac{1}{\sum A_i} \sum_{i: A_i = 1} Y_i -  \frac{1}{\sum A_i} \sum_{i: A_i = 1} \hat{\mathbb{E}}[Y_i \mid A = 0, M_i, C_i],$$
and estimate the organic indirect effect as
$$\hat{\mathbb{E}}[Y^{(0, I = 1)}] - \hat{\mathbb{E}}[Y_0] = \frac{1}{\sum A_i} \sum_{i: A_i = 1} \hat{\mathbb{E}}[Y_i \mid A = 0, M_i, C_i] - \frac{1}{\sum(1 - A_i)} \sum_{i: A_i = 0} Y_i.$$
\end{enumerate}
When the data is fully observed, this semi-parametric estimator reduces the necessary modeling assumptions by only requiring outcome model specification under $A = 0$. If values of the mediator are informatively missing, e.g. because the mediator measurements are subject to an assay lower limit, an additional model is specified for the mediator. 

\section{Semiparametric estimators for the organic direct and indirect effects relative to a = 0 for mediators with assay limits}\label{sec:semip_al}

If $M$ is completely observed, the semiparametric procedure from Section \ref{sec:semip} specifies an outcome model only. When $M$ is left censored by an assay lower limit, the missing mediator information can be supplemented by distributional assumptions and parametric modeling. Several existing missing data estimation procedures can be applied to jointly estimate the model parameters of $M$ and $Y.$ The three-step estimation approach introduced in Section \ref{sec:semip} can be extended to mediators with an assay lower limit as follows. Let $\delta$ be an indicator of whether a mediator value $M$ is above the assay lower limit $AL$, i.e. $\delta = 1$ if $M > AL$ and $\delta = 0$ if $M \leq AL.$

\begin{enumerate}
    \item Jointly fit the following generalized linear models with link functions $h(\cdot)$ and $g(\cdot)$ on treated and untreated subjects respectively:
    \begin{align*}
        h(\mathbb{E}[M_i \mid A = 1, C_i]) &= \alpha_0 + \alpha_1^\top C_i, \\
        g(\mathbb{E}[Y_i \mid A = 0, M_i, C_i]) &= \beta_0 + \beta_2 M_i + \beta_3^\top C_i.
    \end{align*}
    We present methods to fit these models in Sections \ref{sec:extra} - \ref{sec:mcem}.
    \item Sample $M_{ij}$ for $j = 1,...,J$ and for those $i = 1,...,N$ with $\delta_i = 0$ from a truncated distribution restricted to $M \leq$ AL. 
    \item For $A_i = 1$, predict $\hat{\mathbb{E}}[Y_i \mid A = 0, M_i, C_i] = \begin{cases} g^{-1}(\hat{\beta}_0 + \hat{\beta}_1 M_i + \hat{\beta}_2 C_i) & \text{if } \delta_i = 1 \\
    \frac{1}{J}\sum_{j = 1}^Jg^{-1}(\hat{\beta}_0 + \hat{\beta}_1 M_{ij} + \hat{\beta}_2 C_i) & \text{if } \delta_i = 0.
    \end{cases} $
    \item Estimate the organic direct effect relative to $a = 0$ by
$$\hat{\mathbb{E}}[Y_1] - \hat{\mathbb{E}}[Y^{(0, I = 1)}] = \frac{1}{\sum A_i} \sum_{i: A_i = 1} Y_i -  \frac{1}{\sum A_i} \sum_{i: A_i = 1} \hat{\mathbb{E}}[Y_i \mid A = 0, M_i, C_i],$$
and estimate the organic indirect effect relative to $a = 0$ by
$$\hat{\mathbb{E}}[Y^{(0, I = 1)}] - \hat{\mathbb{E}}[Y_0] = \frac{1}{\sum A_i} \sum_{i: A_i = 1} \hat{E}[Y_i \mid A = 0, M_i, C_i] - \frac{1}{\sum(1 - A_i)} \sum_{i: A_i = 0} Y_i.$$
\end{enumerate}

For Step 1 of the algorithm, approaches to parameter estimation with left censoring can be roughly classified into procedures that directly maximize an observed data likelihood with numerical optimization procedures and procedures that maximize a function of the full data likelihood. 

The methods described in Sections \ref{sec:semip} and \ref{sec:semip_al} are generalizable to any observed $C$, $M$, and $Y$. Based on the motivating example of the HIV curative treatment, consider a continuous mediator $M$ measured with an assay lower limit $AL$ and a binary outcome $Y$ for a sample of size $N$. The pre-treatment common causes of the mediator and outcome are collected in a vector $C$ and can be of any type. For $i = 1,...,N$ let $\tilde{M}_i$ be the true mediator values. We assume that given $C_i = c_i$ they follow a normal distribution; specifically, we assume  $\tilde{M}_i \mid C_i = c_i \sim N(\alpha_0 + \alpha_1^\top c_i, \sigma^2_M)$. Let $M_i$ be the observed mediator values with $\tilde{M}_i = M_i$ for values measured above the assay lower limit. Further assume a logistic regression model for $Y_i$; that is $Y_i \mid C_i = c_i, A_i = 0, \tilde{M}_i = m_i \hspace{0.5em} \sim \hspace{0.5em} Binomial(p_{\beta}(m_i, c_i))$ where $p_{\beta}(m_i, c_i)$ is related to $M$ and $C$ through a logit link function i.e. $h\large(p_{\beta}(m_i, c_i)\large) = \log\Big(\frac{p_{\beta}(m_i, c_i)}{1- p_{\beta}(m_i, c_i)}\Big)$ and $$p_{\beta}(m_i, c_i) = \frac{\exp(\beta_0 +\beta_1 m_i + \beta_2^\top c_i)}{1 + \exp(\beta_0 +\beta_1 m_i + \beta_2^\top c_i)}.$$ The full data is $F_i = (C_i, A_i = 0, \tilde{M}_i, Y_i)$ for $i = 1,...,N$ and the observed data is $O_i = (C_i, M_i, \delta_i, Y_i)$ for $i = 1,...,N.$ 

\subsection{Extrapolation method}\label{sec:extra}

If we consider estimation of the outcome and mediator models separately, the problem simplifies to well-known results with consistent estimates. The outcome model can be constructed differently for values of $M$ above and below the assay limit, since the model conditions on $M$; thus one could fit a logistic regression model for $Y$ when observations of $M$ are above the assay limit. To estimate the parameters of the mediator model in step one of the algorithm we apply the maximum likelihood approach described in \cite{aitkin1981}. The observed data likelihood of the mediator distribution is given by     
\begin{align*}
    L(\vec{\alpha}, \sigma^2_M) &= \prod_{i = 1}^n f(m_i \mid c_i)^{\delta_i}[P(M_i \leq AL \mid C_i = c_i)]^{1-\delta_i} \\
        &= \prod_{i = 1}^n \Bigg[ \frac{1}{\sqrt{2\pi\sigma^2_M}} \exp \Bigg(- \frac{(m_i - \vec{\alpha}^\top c_i)^2}{2\sigma^2_M}\Bigg) \Bigg]^{\delta_i}\Bigg[\Phi\Bigg(\frac{AL - \vec{\alpha}^\top c_i}{\sigma_M}\Bigg) \Bigg]^{1-\delta_i},
\end{align*}
where $\Phi$ is the standard normal cumulative distribution function.

The log likelihood is
\begin{align*}
    l(\vec{\alpha}, \sigma^2_M) &= \sum_{i = 1}^n \delta_i \Bigg(-\frac{1}{2}\log(2\pi \sigma^2_M)  - \frac{(m_i - \vec{\alpha}^\top c_i)^2}{2\sigma^2_M}\Bigg) + (1-\delta_i) \Bigg( \log  \Phi\Bigg(\frac{AL - \vec{\alpha}^\top c_i}{\sigma_M}\Bigg) \Bigg).
\end{align*}
McCulloch \cite{mcculloch1997} derived an iterative maximum likelihood (ML) procedure to estimate $(\alpha_0, \alpha_1, \sigma^2_M$). After initializing parameter values at $t = 0$, the procedure updates $\sigma^{2(t+1)}_M$ using data and prior parameter values ($\alpha^{(t)}$, $\sigma^{2(t)}_M$):   
$$\sigma^{2(t+1)}_M = \frac{\sum_{i = 1}^n\delta_i(m_i - {\alpha^{(t)}}^{\top}  c_i)^2}{\sum_{i = 1}^n\Bigg[\delta_i  + (1-\delta_i) \frac{\phi\Big(\frac{AL - ^{(t)} c_i}{\sigma_M^{(t)}}\Big)\frac{AL - {\alpha^{(t)}}^{\top} c_i}{\sigma_M^{(t)}}} {\Phi\Big(\frac{AL - {\alpha^{(t)}}^{\top} c_i}{\sigma_M^{(t)}}\Big)}\Bigg]},$$ where $\phi(\cdot)$ is the density of the standard normal distribution. Then, the $\alpha$'s are updated by solving the score equation with respect to the $\alpha$'s:
\begin{equation}\label{eqn:normeqncen}
    \frac{\partial l}{\partial \vec{\alpha}} = \sum_{i = 1}^n\delta_i\frac{(m_i - \vec{\alpha}^{\top} c_i) c_i^\top}{\sigma^2_M} + (1-\delta_i) \frac{\Bigg(\vec{\alpha}^{\top} c_i - \sigma_M\phi\Big(\frac{AL - \vec{\alpha}^\top c_i}{\sigma_M}\Big)\Big/\Phi\Big(\frac{AL - \vec{\alpha}^\top c_i}{\sigma_M}\Big) - \vec{\alpha}^{\top} c_i\Bigg) c_i^\top}{\sigma^2_M}, 
\end{equation}
Estimation of $\alpha$ can be simplified by recognizing the familiar form of the score function that resembles the normal equations given by least squares estimation with left censored observations replaced by
\begin{equation}\label{eqn:normimp}
   m_i^{(t)} = {\vec{\alpha}^{\top(t)}}\mathbf{c_i} - \frac{\sigma_M^{(t)}\phi\Big(\frac{AL - {\vec{\alpha}^{\top(t)}} c_i}{\sigma_M^{(t)}}\Big)}{\Phi\Big(\frac{AL - {\vec{\alpha}^{\top(t)}} c_i}{\sigma_M^{(t)}}\Big)}. 
\end{equation}
The "normal equations" (\ref{eqn:normeqncen}) suggest a simple iterative imputation procedure for estimating the coefficients of the mediator model. After imputing the $m_i^{(t)}$ using (\ref{eqn:normimp}), a simple OLS model is fit to update the estimates of $\alpha.$ The algorithm iterates between updating $\sigma^2_M$ and $\alpha$ until a convergence criterion is met. Details of the derivations are presented in Web-appendix \ref{app:extra}.

The advantage of the extrapolation method is its speed. However, since the outcome model is only fit on untreated subjects with mediator values above the assay lower limit, the sample used to fit the outcome model can be quite small leading to large standard errors. Furthermore, the outcome model extrapolates for mediator values below the assay limit.

\subsection{Numerical optimization and integration}\label{sec:noptim}

Cole \cite{cole2009estimating} estimated the odds of HIV treatment naivete as a function of HIV viral load measured with a lower detection limit by numerical integration and optimization of the observed data log likelihood. We extend this approach to estimate causal indirect and direct effects. The observed data likelihood is given by:
\begin{align*}
L &= \prod_{i = 1}^N \Bigg[f(y_i \mid m_i, c_i; \vec{\beta})f(m_i \mid c_i; \vec{\alpha},  \sigma_M^2)\Bigg]^{\delta_i}\times \Bigg[\int_{-\infty}^{AL}f(y_i \mid m, c_i; \vec{\beta})f(m_i \mid c_i; \vec{\alpha},  \sigma_M^2)dm\Bigg]^{1-\delta_i}. 
\end{align*}  
Under the data generating assumptions in Section \ref{sec:semip_al}, the log likelihood $l$ is
\begin{align*}
l &= \sum_{i = 1}^{N} \delta_{i} \Bigg[y_i(\beta_0 + \beta_1 m_i + \beta_2^\top c_i)- \log(1+\exp(\beta_0 + \beta_1 m_i + \beta_2^\top c_i))-\frac{1}{2}\log\Big(2\pi\sigma^2_M\Big) - \frac{(m_i - \vec{\alpha}^\top c_i)^2}{2\sigma^2_M} \Bigg] \\
&\quad + (1 - \delta_i) \int_{-\infty}^{AL}\Bigg[y_i (\beta_0 + \beta_1 m + \beta_2^\top c_i)- \log(1+\exp(\beta_0 + \beta_1 m + \beta_2^\top c_i)) -\frac{1}{2}\log\Big(2\pi\sigma^{2}_M\Big) - \frac{(m - \vec{\alpha}^{\top} c_i)^2}{2\sigma^{2}_M} \Bigg]dm.
\end{align*}

The R programming language has efficient built-in functions for numerical integration and optimization \texttt{integrate()} and \texttt{optim()}  to accommodate the integrated density function and non-linearity in $M$. We constrained the optimization space for $\sigma^2_M$, the variance of the mediator distribution, to ensure that $\sigma^2_M$ non-negative.

\subsection{Monte Carlo EM}\label{sec:mcem}

The methodology of the Expectation-Maximization (EM) algorithm \cite{dempster1977} for maximum likelihood estimation with a latent variable is an alternative estimation approach to the assay limit problem. As we saw in Section \ref{sec:noptim}, the observed data likelihood involves complex optimization and numerical integration over the censored values. On the other hand, the full data likelihood often has a familiar form, especially if the outcome is in the exponential family. If the assay did not have a lower limit we would have the full data $(Y, \tilde{M}, C)$ and likelihood function  
\begin{align*}
\tilde{L} &= \prod_{i = 1}^N f(y_i \mid \tilde{m}_i, c_i, A = 0; \vec{\beta})f(\tilde{m}_i \mid c_i; \vec{\alpha},  \sigma_M^2). 
\end{align*}
Because of the assay limit, instead of observing $\tilde{M}$ we observe the values of $\tilde{M}$ only if they are above the assay limit. If the censored mediator observations, those with $\delta_i = 0$, are viewed as latent random variables, maximum likelihood estimation can proceed with the EM algorithm. Let $\theta$ be a vector of the $\alpha$'s, $\sigma^2_M$, and $\beta$'s. The algorithm begins with initial values $\theta^{(0)}$ and iterates between the E and M step until the absolute difference between $\theta^{(t)}$ the current estimate and $\theta^{(t+1)}$ the updated estimate is small. The E (Expectation) step, replaces the log likelihood with the expectation of the full log likelihood conditional on current parameter values $\theta^{(t)}$ and observed data $(Y_i, M_i,\delta_i, C_i), \quad i = 1,..., N$:
\begin{align*}
  \mathbb{Q}(\theta \mid \theta^{(t)}) &= \mathbb{E}\Big[\tilde{l}(\theta \mid \tilde{M}_i, Y_i, C_i) \mid Y_i = y_i, M_i = m_i, C_i = c_i; \theta^{(t)} \Big] \\ 
    &= \sum_{i = 1}^N\delta_i \Big(\log f(y_i \mid m_i, c_i; \theta) + \log f(m_i \mid c_i; \theta) \Big) \\ &\qquad + (1-\delta_i)\mathbb{E}\Big[\log f(Y_i \mid \tilde{M}_i, c_i; \theta)  
    + \log f(\tilde{M}_i \mid C_i; \theta) \mid Y_i = y_i, M_i = m_i, C_i = c_i; \theta^{(t)}\Big].
\end{align*}
The M (Maximization) step, updates $\theta^{(t)}$ as the value of $\theta$ which maximizes $\mathbb{Q}(\theta \mid \theta^{(t)})$, i.e. $$\theta^{(t+1)} = \argmax_\theta \mathbb{Q}(\theta \mid \theta^{(t)}).$$ The EM algorithm iterates between the E and M steps until a convergence criterion is met. The EM algorithm provides a maximum likelihood approach for missing data, but depending on the type of $M$ and $Y$ further assumptions and modifications are necessary. In our setting, a binary $Y$ and a continuous $M$ present the following two challenges: 1. $\mathbb{Q}(\theta \mid \theta^{(t)})$ requires knowledge of the  distribution $\tilde{M}$ given $Y$ and 2. the assumptions on the distribution of the binary outcome leads to $\mathbb{Q}(\theta \mid \theta^{(t)})$ with a non-linearity in $M$.

Wei et al. \cite{wei1990} suggest an adaptation to the EM algorithm known as the Monte Carlo EM (MCEM) algorithm, which estimates $\mathbb{Q}(\theta \mid \theta^{(t)})$ with a Monte Carlo (MC) estimate with samples taken from the distribution of $M$ given $Y$.

\subsubsection{Monte Carlo EM Algorithm (MCEM): modified E-step}

The MCEM algorithm approximates the E-step of the EM algorithm with an MC estimate, simplifying the integration of non-linear functions of $M.$ For the data model from Section \ref{sec:semip_al}, $\mathbb{Q}(\theta \mid \theta^{(t)})$ is estimated by: 
\begin{align*}
\hat{\mathbb{Q}}(\theta \mid \theta^{(t)}) &= \sum_{i = 1}^{N} \delta_{i} \Bigg[y_i(\beta_0 + \beta_1 m_i + \beta_2 c_i)- \log(1+\exp(\beta_0 + \beta_1 m_i + \beta_2 c_i))-\frac{1}{2}\log\Big(2\pi\sigma^2_M\Big) - \frac{(m_i - \vec{\alpha}^\top c_i)^2}{2\sigma^2_M} \Bigg] \\
&\quad + \frac{1 - \delta_i}{J} \sum_{j = 1}^J\Bigg[y_i (\beta_0 + \beta_1 m_{ij}^{(t)}+ \beta_2 c_i)- \log(1+\exp(\beta_0 + \beta_1 m_{ij}^{(t)} + \beta_2 c_i)) -\frac{1}{2}\log\Big(2\pi\sigma^{2}_M\Big) - \frac{(m_{ij}^{(t)} - \vec{\alpha}^{\top} c_i)^2}{2\sigma^{2}_M} \Bigg],
\end{align*}
where the $m_{ij}^{(t)}$ are sampled from the distribution of $M_i$ given $Y_i, C_i, \theta^{(t)}$ and $M_i$ below the assay limit. The distribution of $M_i$ given $Y_i, C_i, \theta^{(t)}$ and $M_i$ below the assay limit does not have a closed form and must be sampled through distributional approximations. There are many valid methods for sampling; we focus on the grid method for its simplicity and ease of implementation. 
From Bayes rule, the density of $M_i$ given $Y_i, C_i, \theta^{(t)}$ and $M_i$ below the assay limit for observations with $\delta_i = 0$ is $$f(m \mid y, c_i, m \leq AL, y; \theta) = \frac{f(y \mid c, m; \theta) f(m \mid c, m \leq AL; \theta)}{\int_{-\infty}^{AL}f(y \mid c, m; \theta) f(m \mid c, m \leq AL; \theta)dm}.$$
Since the mediator is one-dimensional we can sample from this distribution using a grid approximation, dividing the domain of $M$ below the assay limit into a grid of $K$ equally spaced values $m^{(1)},...,m^{(K)}$. Then for $k = 1,...,K$ evaluate: $$\hat{f}(m^{(k)} \mid c, m \leq AL, y; \theta) = \frac{\hat{f}(y \mid c, m^{(k)}, \theta) \hat{f}(m^{(k)} \mid c, m \leq AL; \theta)}{\sum_{r = -\infty}^{AL}\hat{f}(y \mid c, m^{(r)}, \theta) \hat{f}(m^{(r)} \mid c, m \leq AL; \theta)}.$$ Finally, sample from the $m^{(1)}, ..., m^{(K)}$ values with weights $\hat{f}(m^{(k)} \mid c, m \leq AL, y; \theta).$

\subsubsection{The M-step}

The M-step involves optimizing the MC estimate with respect to $\sigma^2_M, \alpha, \beta$. The updates for the mediator distribution parameters are
\begin{equation*}
  \vec{\alpha}^{\top(t + 1)} = \Bigg(\sum_{i = 1}^N  c_i c_i^\top \Bigg)^{-1}\Bigg(\sum_{i = 1}^N\delta_i m_i c_i^\top + (1-\delta_i)\frac{1}{J}\sum_{j = 1}^J m_{ij}^{(t)} c_i^\top\Bigg)^\top,
\end{equation*}
and
\begin{equation*}
    \sigma^{2(t+1)}_{M} = \frac{1}{N}\sum_{i = 1}^{N}\Bigg(\delta_i(m_i - \vec{\alpha}^{\top(t + 1)}c_i)^2  + \frac{1-\delta_i}{J} \sum_{j = 1}^J(m_{ij} - \vec{\alpha}^{\top(t + 1)} c_i)^2\Bigg).
\end{equation*}
The outcome model leads to the score function:
\begin{equation*}
    \frac{\partial \mathbb{Q}(\theta \mid \theta^{(t)})}{\partial\vec{\beta}} = \sum_{i = 1}^{N}\Bigg(\delta_i \begin{bmatrix} 1 \\ m_i \\ c_i \end{bmatrix}
    \Big(y_i - p_{\beta}(m_i, c_i)\Big) + (1-\delta_i) \frac{1}{J} \sum_{j = 1}^J \begin{bmatrix} 1  \\ m_{ij}^{(t)} \\ c_i \end{bmatrix} \Big(y_i - p_{\beta}(m_{ij}^{(t)}, c_i)\Big)\Bigg).
\end{equation*} 
These score functions are almost identical to those of the classical generalized linear model setting. The terms with mediator values below the assay limit are weighted by the inverse number of samples taken from the mediator distributions. Thus, standard statistical software can be applied to implement the M-step with weights equal to $1$ for uncensored observations and to $1/J$ for censored observations. 

\section{Simulation study}\label{sec:sim}

This simulation study evaluates the proposed estimators for the indirect effect of $A$ on $Y$ relative to $a = 0$ through a mediator $M$ that is subject to an assay lower limit. The simulated data closely follow the estimated data distribution of the HIV cure data analyzed in Section \ref{sec:dataapp} and the assumed data models from Section \ref{sec:semip_al}. The estimation target of Section \ref{sec:dataapp} is the indirect effect of a hypothetical treatment $A$ on viral suppression through week 4, $Y$, mediated by two viral reservoir measures $M$. The simulation focuses one of the measured viral reservoirs, cell-associated HIV RNA. We simulate a single binary pre-treatment common cause of the mediator and outcome, $C$. The hypothetical treatment $A = 1$ is assumed to cause a shift in the mediator distribution. Appendix \ref{app:shift} provides more details on estimating the indirect effect of a hypothetical treatment assumed to shift the distribution of the mediator. Under $A = 0$, the distribution of $M$, the log$_{10}$ viral reservoir, follows a normal distribution with mean $\alpha_0 + \alpha_1 C$ and variance $ \sigma^2_M$ with $(\alpha_0, \alpha_1) = (2.03, 0.14)$ and $\sigma_M^2 = 0.78$. The distribution of $Y$, viral rebound through week 4, is Bernoulli with probability of $Y  = 1$ equal to $p_\beta(m,c) = \frac{\exp(\beta_0 + \beta_1 m + \beta_2 c)}{1+\exp(\beta_0 + \beta_1 m + \beta_2 c)}$ with $(\beta_0, \beta_1, \beta_2) = (0.84, -0.73, 1.39).$ The simulated scenarios evaluate the effect of different sample sizes and mediator shifts on the estimators. A moderate sample (N = 100) resembles our data example, and a large sample (N = 500) reflects large sample properties. Mediator shifts reflect a range of biologically significant effects of a hypothetical curative intervention: 0.5, 1, and 2 \cite{hill2014}. 

The simulation study evaluates the bias and variance of the proposed methods from Section \ref{sec:semip_al}: 1. mediator - outcome extrapolation (Section \ref{sec:extra}) 2. numerical optimization of the joint outcome-mediator log likelihood  (Section \ref{sec:noptim}) 3. Monte Carlo EM (Section \ref{sec:mcem})  4. imputing censored mediator values with half the assay lower limit. 

\begin{figure}[h!]
\captionsetup{width=\textwidth}
\caption{Simulations of methods for estimating the indirect effect of $A$ on $Y$ relative to $a = 0$ mediated by $M$ with a lower assay limit of 1.96 as in the HIV cure application. The methods considered are: 1. mediator - outcome extrapolation (Section \ref{sec:extra}), 2. numerical optimization of the joint outcome-mediator log likelihood (Section \ref{sec:noptim}), 3. Monte Carlo EM (Section \ref{sec:mcem}), and 4. imputing censored mediator values with half the assay lower limit. The rows depict sample size of $N = 100$ or $N = 500$; columns display mediator shifts of 0.5, 1, and 2. The true indirect effects 0.07, 0.15, and 0.28 for shifts of 0.5, 1, and 2, respectively, are displayed as dashed lines.}
  \includegraphics[width = \linewidth]{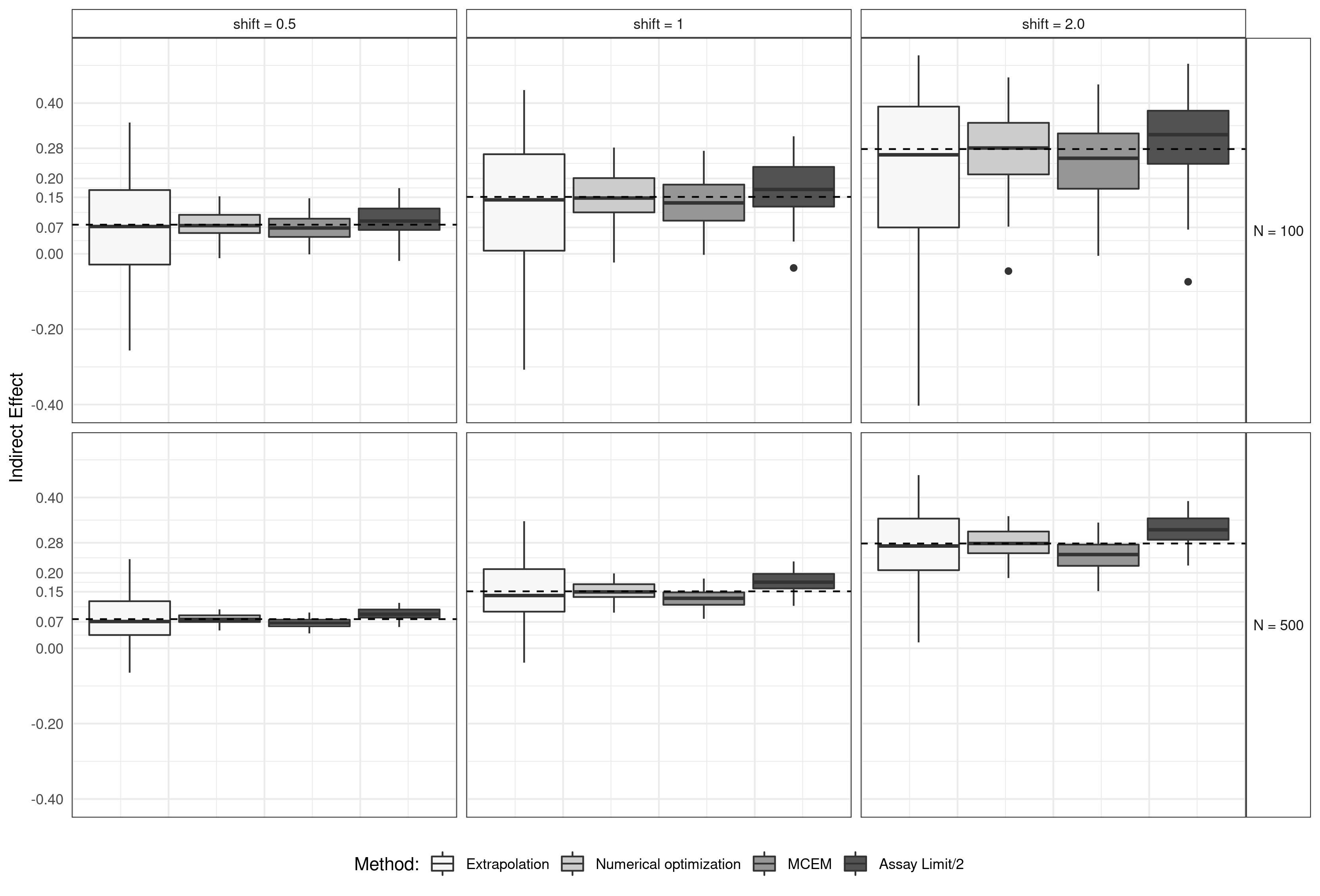}
  \label{fig:sim}
\end{figure}

Figure \ref{fig:sim} shows the influence of sample size and mediator shift size on the performance of the estimators. Larger sample sizes and smaller mediator shifts are associated with estimators with lower variance. Since all estimation procedures involve model extrapolation, a larger mediator shift is associated with extrapolations further outside of the observed range of the data. For every evaluated sample size and mediator shift, the method with the lowest bias and variance is numerical optimization. The extrapolation method has the highest variance, attributed to fitting the outcome model on a limited sample of mediator values above the assay limit. The Monte Carlo EM estimator improves on the bias of the naive imputation of half the assay limit but does not achieve the low bias of numerical optimization. 

\section{The indirect effect of HIV curative treatments that shift the distribution of the viral reservoir}\label{sec:dataapp}

Antiretroviral therapy (ART), the current standard of care, successfully suppresses the HIV viral load in the blood, but ART is not a cure and requires lifelong use \cite{who2019}. Additionally, ART use is associated with side effects, drug-drug interactions, and drug resistance, and ART may result in pill fatigue from daily use. A treatment that targets the elimination of the HIV reservoir aims to replace ART and achieve HIV remission \cite{pitman2018}. Testing new curative treatments requires an analytic treatment interruption (ATI) study, which disrupts ART until the time of viral rebound, when the viral load reaches a pre-defined threshold. The risk to participants and the cost of conducting an ATI study limits opportunities for large clinical trials. Using ATI study data, Li et al. \cite{li2016} found that biomarkers of a larger on-ART viral reservoir were associated with a shorter off-ART time to viral rebound. 

\begin{figure}[h]
\captionsetup{width=0.8\textwidth}
\caption{\small Causal diagram for the effect of an HIV curative treatment that shifts the distribution of the pre-ATI HIV cell-associated RNA}
\begin{center}
{\begin{tikzpicture}[%
	->,
	>=stealth,
	node distance=2cm,
	pil/.style={
		->,
		thick,
		shorten =6 pt,}
	]
	\node (1) {M: viral reservoir};
	\node[left=of 1] (2) {$A:$\begin{tabular}{l}
	     HIV curative \\
	     treatment
	\end{tabular}};
	\node[right=of 1] (3) {$Y:$\begin{tabular}{l}
	     viral suppression \\
	     through week 4 \\
	     after ART \\ 
	     interruption
	\end{tabular}};
	\node[below= of 1] (4) {$C$: NNRTI-based};
	\draw [->] (1.east) -- (3.west);
	\draw [->] (2) to [out=30, in=150] (3);
	\draw [->] (2.east) to (1);
	\draw [->] (4) to (1);
	\draw [->] (4) to (3);
\end{tikzpicture}}
\end{center}
\label{fig:hivdag}
\end{figure}
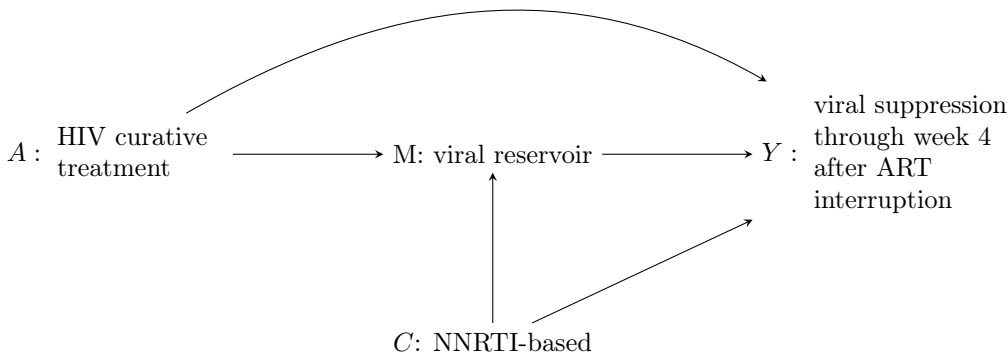

The causal diagram in Figure \ref{fig:hivdag} is a visual representation of the HIV curative treatment mediation question. For viral suppression through week 4 after ATI, the relevant common cause is the type of initial HIV treatment regimen: NNRTI-based ART versus non-NNRTI-based ART. The indirect effect can inform the reduction in the viral reservoir necessary to meaningfully increase the probability of viral suppression through week 4. Lok and Bosch \cite{lok2021} estimated the organic indirect effect relative to $a = 0$ on viral rebound $Y$ following ATI of an HIV curative treatment $A$ that would reduce the size of the viral reservoir as measured by two HIV persistence measures $M$.  The organic and pure indirect effect relative to $a = 0$ can be estimated without outcome data under treatment.  

Measures of HIV viral persistence are subject to an assay limit. Lok and Bosch\cite{lok2021} addressed the assay limit by assuming that viral rebound is not associated with how far mediator values fall below the assay lower limit. We re-analyzed the data with the methods described in Section \ref{sec:semip_al}, which assume varying effects on viral rebound based on how far the viral persistence measure, cell-associated RNA (CA-RNA) or single copy RNA (SCA-RNA), falls below the assay limit. We assume that CA-RNA and SCA-RNA given the HIV treatment regimen, NNRTI-based (yes or no), are normally distributed on the log scale with a linear relationship of the mean and $C$ and a constant variance. We modelled the binary outcome $Y$, viral suppression through week 4, with logistic regression, assuming the same model for values below and above the assay limit. We consider the indirect effect of treatments that shift the mediator distributions by $0.5\log_{10}$, $1\log_{10}$, $1.5\log_{10}$, $2\log_{10}$ given C. Table \ref{tab:hiv_ie} presents the indirect effects along with non-parametric bootstrap 95\% confidence intervals.

\begin{table}[h!]
\topcaption{Organic indirect effects of curative HIV-treatments that shift the distribution of HIV persistence measures downwards. The estimated probability of no viral rebound through week 4 without curative treatment was 63/124 or 51\%. \\
}
\begin{tabular}{lllcl}
  \toprule
HIV persistence measure & Mediator shift$^c$ & Estimation method$^d$ & Indirect effect & 95\% CI$^e$ \\ 
  \hline
CA HIV-RNA$^a$  & 0.50 & AL / 2 & 6.40\%  &  (2.0\%, 10.2\%) \\ 
 &  & extrapolation & 2.90\%  &  (-2.03\%, 21.7\%) \\ 
 &  & MCEM & 6.80\%  &  (2.20\%, 11.2\%) \\ 
 &  & numerical optimization & 7.70\%  &  (2.70\%, 12.5\%) \\ 
 & 1.00 & AL / 2 & 12.6\%  &  (4.0\%, 19.8\%) \\ 
 &  & extrapolation & 8.7\%  &  (-23.7\%, 32.1\%) \\ 
  & & MCEM & 13.3\%  &  (4.4\%, 21.5\%) \\ 
  & & numerical optimization & 15.1\%  &  (5.4\%, 23.6\%) \\ 
 & 1.50 & AL / 2 & 18.4\%  &  (5.9\%, 28.0\%) \\ 
  & & extrapolation & 14.3\%  &  (-27\%, 39.4\%) \\ 
  & & MCEM & 19.4\%  &  (6.6\%, 30.1\%) \\ 
 &  & numerical optimization & 21.9\%  &  (8.1\%, 32.7\%) \\ 
 & 2.00 & AL / 2 & 23.7\%  &  (7.9\%, 35.0\%) \\ 
 &  & extrapolation & 19.5\%  &  (-30.6\%, 44.9\%) \\ 
  & & MCEM & 24.9\%  &  (8.7\%, 37.2\%) \\ 
 &  & numerical optimization & 27.8\%  &  (10.8\%, 39.6\%) \\ 
SCA HIV-RNA$^b$  & 0.50 & extrapolation & -21.7\%  &  (-53.6\%, 27.4\%) \\ 
   &  & numerical optimization & 4\%  &  (-1.3\%, 8.6\%) \\ 
   &  & MCEM & 4.4\%  &  (-1.9\%, 7.4\%) \\ 
   &  & AL / 2 & 5.1\% &  (-2.2\%, 10.9\%) \\ 
   & 1.00 & extrapolation & -25\%  &  (-58\%, 33.5\%) \\ 
   &  & numerical optimization & 7.8\%  &  (-2.7\%, 16.6\%) \\ 
   &  & MCEM & 8.6\%  &  (-3.9\%, 14.2\%)  \\ 
   &  & AL / 2 & 9.9\% &  (-4.4\%, 20.5\%) \\ 
   & 1.50 & extrapolation & -28.1\%  &  (-60\%, 38.3\%) \\ 
   &  & numerical optimization & 11.6\%  &  (-4\%, 23.7\%) \\ 
   &  & MCEM & 12.6\%  &  (-5.9\%, 20.7\%) \\ 
   &  & AL / 2 & 14.6\%  &  (-6.6\%, 28.5\%) \\ 
   & 2.00 & extrapolation & -31\%  &  (-61\%, 41.7\%) \\ 
   &  & numerical optimization & 15.1\%  &  (-5.4\%, 30.1\%) \\ 
   &  & MCEM & 16.4\%  &  (-8\%, 26.2\%) \\ 
   &  & AL / 2 & 18.8\%  &  (-8.7\%, 35\%) \\ 
   \bottomrule
\end{tabular}
\label{tab:hiv_ie}
{\parbox{6.2in}{
\footnotesize \raggedright $^a$ Cell-Associated HIV-RNA, on-ART (N = 124). \\
 $^b$ Single Copy HIV-RNA, on-ART (N = 124). \\
$^c$ Treatment-induced downward shift of the viral persistence distribution on the $log_{10}$ scale. \\
$^d$ Extrapolation: mediator - outcome extrapolation (Section \ref{sec:extra}), \\ 
\quad numerical optimization: numerical optimization of the joint outcome-mediator log likelihood (Section \ref{sec:noptim})\\ 
\quad MCEM: Monte Carlo EM (Section \ref{sec:mcem}), \\ 
\quad AL/2: imputing censored mediator values with half the assay lower limit.  \\
$^e$ Risk difference scale (in percent). Non-parametric bootstrap 95\% confidence intervals. \\}}
\end{table}

Since the numerical optimization approach performed best in the simulation studies, we will focus on those estimates. A $0.5\log_{10}$ downward shift in CA HIV-RNA is associated with a 7.7\% (95\% CI: 2.7\%, 12.5\%) absolute reduction in the probability of week 4 viral suppression after ATI. At the other extreme, a 100-fold reduction (equivalent to a $2\log_{10}$ downward shift) in CA-RNA is associated with a 27.8\% (95\% CI: 10.8\%, 39.6\%) reduction in risk of viral rebound through week 4. Thus, a sizeable mediator shift is necessary to substantially reduce the risk of viral rebound through week 4 after ATI, consistent with mathematical modeling \cite{hill2014}. In contrast, a 3-fold reduction in SCA-RNA (equivalent to a $0.5\log_{10}$ downward shift) is associated with a 4\% (95\% CI: -1.3\%, 8.6\%) reduction in probability of week 4 viral suppression after ATI and a 100-fold reduction (equivalent to a $2\log_{10}$ downward shift) in SCA-RNA is associated with a 15.1\% (95\% CI: -5.4\%, 30.1\%) reduction in risk of viral rebound through week 4.

Lok and Bosch \cite{lok2021} addressed a similar HIV curative question and found a similar reduction in risk after a 3-fold reduction of viral reservoir, 6.9\% (95\% CI: 1.7\%, 12.6\%). However, the maximum achievable risk reduction was estimated to be 12.7\% (3.0\%,22.7\%), consistent with a treatment that shifts all mediator values below the assay limit. The difference between our findings and those in \cite{lok2021} are attributable to the difference in assumptions. \cite{lok2021} assumed that how far viral reservoir values fall below the assay limit was not associated with viral rebound through week 4. Thus, the maximal shift that can be considered is a shift which moves all values below the assay limit. On the other hand, our methods assume that the probability of viral suppression depends on how far below the assay limit a mediator values lies, and we can consider larger treatment-induced shifts in the viral reservoir. 

\section{Discussion}

Mediation analysis is  a key methodological tool in the evaluation of new therapies, but when the mediator is a biomarker subject to an assay limit, additional estimation tools are required. Our semi-parametric estimates of the organic indirect and direct effects build on Lok and Bosch \cite{lok2021}, which addressed the assay limit problem by assuming that the outcome is not affected by how far the mediator values fall below the assay limit. We assume that the outcome is associated with how far the values fall below the assay limit, and make parametric assumptions on the mediator distribution and the expected outcome for mediator values above and below the assay limit. We developed and evaluated three methods: 1. extrapolation 2. numerical integration and optimization of the observed data likelihood function and  3. the Monte Carlo Expectation-Maximization (MCEM) algorithm.

A simulation study designed to emulate the HIV cure application compared the proposed methods to address the assay limit problem. Simulation scenarios included different sample sizes and mediator shifts. Numerical optimization of the observed likelihood performed best in all our simulation scenarios. Large mediator shifts can be evaluated using our methods, but the larger the mediator shift the larger the variability in the estimate, attributable to extending the model further beyond the observed mediator values. 

We applied our methods to an HIV curative example that aims to use existing ART interruption data without curative treatment to evaluate the required treatment-induced shift necessary to substantially increase the probability of viral suppression through week 4. For a $2\log_{10}$ downward shift of CA-RNA, the largest HIV viral persistence shift considered, the indirect effect estimated by numerical optimization and integration resulted in an indirect effect of $27.8\%$ $(10.8\%, 39.6\%)$. For an equivalent shift of $2\log_{10}$ in the SCA-RNA, the indirect effect was $15.1\%$ $(-5.4\%, 30.1\%)$. Thus, targeting CA-RNA to delay viral rebound might be more effective treatment of HIV than targeting SCA-RNA.

Hill et al. \cite{hill2014} used mathematical models to explore the HIV viral reservoir reduction necessary to substantially extend the time to viral rebound. They found that a roughly 2000-fold reduction results in a median extension of the time to viral rebound by about a year. The authors suggest using their models to ascertain uncertainty induced by the assay lower limit of viral reservoir measures. In contrast, we suggest direct estimation methods that incorporate values below the assay limit into the model. Future research could combine our methods with the models introduced by \cite{hill2014} to attain a better understanding of the viral reservoir reduction necessary for a substantial extension of time to viral rebound.   

Assay limits are not unique to HIV research. Our methods could be useful in the area of environmental exposures with devices constrained by detection limits. For example, air-borne transmission of SARS CoV-2 is an established mode of viral infection \cite{who2020}. Estimating the effect of a preventative measure $A$ such as mask wearing on COVID-19 infection $Y$ mediated by the quantity of aerosolized SARS CoV-2 $M$ could provide insight into effective prevention of SAR CoV-2 infection. Airborne viral particles are most commonly measured by quantitative reverse transcription polymerase chain reaction (qRT-PCR) subject to a lower limit of detection \cite{kim2020}. Occupational exposure is another example of an active research area with a limit of detection. For example, in dental prosthetic laboratories with manual production, particulate matter is released when abrasion and polishing are performed. Particulate matter has been reported to directly cause lung cancer as well as other respiratory illnesses \cite{yildirim2020}. The effect of an intervention $A$ aimed at reducing respiratory disease incidence $Y$ by limiting (or mediated by) occupational exposure to particulate matter $M$ could lead to better methods of safe-guarding the lab technicians. Particulate matter $M$ can be measured using an inductively coupled plasma mass spectrometer (ICP-MS) device that can only measure particulate matter above a limit of detection \cite{yildirim2020}.

We focused on mediation analysis with a continuous mediator and a binary outcome, but an advantage of the described methods is their adaptability to other outcome types. For example, if the outcome of interest is also a continuous biomarker, the methods for estimating the organic indirect and direct effects are easily modified by replacing the logistic regression model with a normality assumption and linear regression. In the curative HIV example, we modelled the outcome as a binary variable of whether there was viral rebound through week 4. Instead, we might focus on the difference in time to viral rebound. Our methods can extend to time-to-event outcomes but will require additional methods and assumptions to handle right or interval censoring. This is an interesting subject for future research. 

\pagebreak

\printbibliography

\pagebreak

\subsection{Appendix Section}

\subsection{Derivations for the extrapolation method}\label{app:extra}

The extrapolation method fits separate models for the mediator and outcome to simplify the estimation in the presence of left censoring. The procedure begins with estimation of the variance $\sigma^2_M$ by differentiating the log likelihood with respect to $\sigma_M$:

\begin{align*}
    \frac{\partial l}{\partial \sigma_M} &= \sum_{i = 1}^n\delta_i \Bigg(-\frac{1}{\sigma_M} + \frac{(m_i - \vec{\alpha}^{\top} c_i)^2}{\sigma_M^3}\Bigg) - (1-\delta_i)\Bigg( \frac{AL - \vec{\alpha}^{\top} c_i}{\sigma_M^2}\Bigg)\frac{\phi\Big(\frac{AL - \vec{\alpha}^\top c_i}{\sigma_M}\Big)} {\Phi\Big(\frac{AL - \vec{\alpha}^\top c_i}{\sigma_M}\Big)} \\
    &= \sum_{i = 1}^n\delta_i \Bigg( \frac{(m_i - \vec{\alpha}^{\top} c_i)^2}{\sigma_M^3}\Bigg) - \delta_i\frac{1}{\sigma_M}  - (1-\delta_i) \frac{1}{\sigma_M}\frac{\phi\Big(\frac{AL - \vec{\alpha}^\top c_i}{\sigma_M}\Big)\frac{AL - \vec{\alpha}^\top c_i}{\sigma_M}} {\Phi\Big(\frac{AL - \vec{\alpha}^\top c_i}{\sigma_M}\Big)} \\
    &= \frac{1}{\sigma_M^3}\Bigg(\sum_{i = 1}^n\delta_i (m_i - \vec{\alpha}^{\top} c_i)^2 - \sigma_M^2\sum_{i = 1}^n \Bigg[\delta_i  + (1-\delta_i) \frac{\phi\Big(\frac{AL - \vec{\alpha}^\top c_i}{\sigma_M}\Big)\frac{AL - \vec{\alpha}^\top c_i}{\sigma_M}} {\Phi\Big(\frac{AL - \vec{\alpha}^\top c_i}{\sigma_M}\Big)}\Bigg]\Bigg). \\
\end{align*}

Solving the score function gives the following iterative estimate:

$$\hat{\sigma}^2_M = \frac{\sum_{i = 1}^n\delta_i(m_i - \alpha^{\top} c_i)^2}{\sum_{i = 1}^n\Bigg[\delta_i  + (1-\delta_i) \frac{\phi\Big(\frac{AL - \vec{\alpha}^\top c_i}{\sigma_M}\Big)\frac{AL - \vec{\alpha}^\top c_i}{\sigma_M}} {\Phi\Big(\frac{AL - \vec{\alpha}^\top c_i}{\sigma_M}\Big)}\Bigg]}.$$

Similarly, we differentiate the log likelihood with respect to $\alpha$

\begin{align*}
    \frac{\partial l}{\partial \vec{\alpha}} &= \sum_{i = 1}^n\delta_i\frac{(m_i - \vec{\alpha}^{\top}c_i)c_i^\top}{\sigma^2_M} - (1-\delta_i) \frac{1}{\sigma_M}\frac{\phi\Big(\frac{AL - \vec{\alpha}^\top c_i}{\sigma_M}\Big)c_i^\top} {\Phi\Big(\frac{AL - \vec{\alpha}^\top c_i}{\sigma_M}\Big)} \\
    &= \sum_{i = 1}^n\delta_i\frac{(m_i - \vec{\alpha}^{\top}c_i)c_i^\top}{\sigma^2_M} - (1-\delta_i) \frac{\sigma_M\phi\Big(\frac{AL - \vec{\alpha}^\top c_i}{\sigma_M}\Big)c_i^\top\Big/\Phi\Big(\frac{AL - \vec{\alpha}^\top c_i}{\sigma_M}\Big)}{\sigma^2_M} - (1-\delta_i)\frac{\vec{\alpha}^{\top}c_ic_i^\top}{\sigma^2_M} + (1-\delta_i)\frac{\vec{\alpha}^{\top}c_ic_i^\top}{\sigma^2_M} \\
    &= \sum_{i = 1}^n\delta_i\frac{(m_i - \vec{\alpha}^{\top}c_i)c_i^\top}{\sigma^2_M} + (1-\delta_i) \frac{\Bigg(\vec{\alpha}^{\top}c_i - \sigma_M\phi\Big(\frac{AL - \vec{\alpha}^\top c_i}{\sigma_M}\Big)\Big/\Phi\Big(\frac{AL - \vec{\alpha}^\top c_i}{\sigma_M}\Big) - \vec{\alpha}^{\top}c_i\Bigg)c_i^\top}{\sigma^2_M}.
\end{align*}

Thus, culminating in a familiar score function similar to the one for linear regression. 

\subsection{Estimating the indirect effect with a mediator shift} \label{app:shift}

The expense and risk to participant safety limit the ability to collect RCT data for HIV curative treatment. Lok and Bosch \cite{lok2021} propose a method to estimate the indirect effect of an HIV curative treatment on viral suppression, mediated by the viral reservoir. The indirect effect is only dependent on $A = 1$ through the distribution of the mediator and if the effect of treatment on the distribution on the mediator is known or hypothesized, we can write the indirect effect as a function of only $A = 0$. For example, assume that treatment reduces the viral reservoir by $\xi$, i.e. $M_1 \sim M_0 - \xi$ :
\begin{align*}
    \mathbb{E}[Y_{I = 1, 0}] - \mathbb{E}[Y_0] &= \int_{m, c} \mathbb{E}[Y\mid M = m, A = 0, C = c]f_{M \mid A=1, C = c}(m) f_C(c) dmdc  - \mathbb{E}[Y_0]\\
        &= \int_{m, c} \mathbb{E}[Y\mid M = m, A = 0, C = c]f_{M \mid A=0, C = c}(m + \xi) f_C(c) dmdc  - \mathbb{E}[Y_0].
\end{align*}
Thus, an estimate for the organic direct effect without treatment data is:
$$\hat{\mathbb{E}}[Y^{(0, I = 1)}] - \hat{\mathbb{E}}[Y_0] = \frac{1}{N} \sum_{i = 1}^N \Big( \hat{E}[Y_i \mid M = m_i - \xi, C = c_i] - Y_i\Big).$$

\end{document}